\documentclass[%
 showpacs,
 showkeys,
 amsmath,amssymb,
]{revtex4-1}

\usepackage{graphicx}
\usepackage{dcolumn}
\usepackage{bm}

\usepackage{CJK}

\usepackage{color}
\usepackage{amssymb}
\usepackage{amsfonts}

\usepackage[colorlinks,urlcolor=blue,linkcolor=blue,citecolor=blue,anchorcolor=blue]{hyperref}

\begin{document}

\preprint{APS/123-QED}

\title{Two-Dimensional Linear and Nonlinear Talbot Effect from Rogue Waves}

\author{Yiqi Zhang$^1$}
\author{Milivoj R. Beli\'c$^{2}$}
\email{milivoj.belic@qatar.tamu.edu}
\author{Milan S. Petrovi\'c$^{2,3}$}
\author{Huaibin Zheng$^1$}
\author{Haixia Chen$^1$}
\author{Changbiao Li$^1$}
\author{Keqing Lu$^4$}
\author{Yanpeng Zhang$^{1}$}
\email{ypzhang@mail.xjtu.edu.cn}
\affiliation{%
 $^1$Key Laboratory for Physical Electronics and Devices of the Ministry of Education \& Shaanxi Key Lab of Information Photonic Technique,
Xi'an Jiaotong University, Xi'an 710049, China \\
$^2$Science Program, Texas A\&M University at Qatar, P.O. Box 23874 Doha, Qatar \\
$^3$Institute of Physics, P.O. Box 68, 11001 Belgrade, Serbia \\
$^4$School of Electronics and Information Engineering, Tianjin Polytechnic University, Tianjin 300387, China
}%

\date{\today}

\begin{abstract}
\noindent
  We introduce two-dimensional (2D) linear and nonlinear Talbot effects.
  They are produced by propagating periodic 2D diffraction patterns and can be visualized as 3D stacks of Talbot carpets.
  The nonlinear Talbot effect originates from 2D rogue waves and forms in a bulk 3D nonlinear medium.
  The recurrences of an input rogue wave are observed at the Talbot length and at the half-Talbot length, with a $\pi$ phase shift;
  no other recurrences are observed.
  Different from the nonlinear Talbot effect, the linear effect displays the usual fractional Talbot images as well.
  We also find that the smaller the period of incident rogue waves,
  the shorter the Talbot length.
  Increasing the beam intensity increases the Talbot length,
  but above a threshold this leads to a catastrophic self-focusing phenomenon
  which destroys the effect.
  We also find that the Talbot recurrence can be viewed as a self-Fourier transform
  of the initial periodic beam that is automatically performed during propagation.
  In particular, linear Talbot effect can be viewed as a fractional self-Fourier transform, whereas the
  nonlinear Talbot effect can be viewed as the regular self-Fourier transform.
  Numerical simulations demonstrate that the rogue wave initial condition is sufficient but not necessary
   for the observation of the effect. It may also be observed from other periodic inputs,
   provided they are set on a finite background.
    The 2D effect may find utility in the production of 3D photonic crystals.
\end{abstract}

\maketitle

%

\section{Introduction}

Recently, optical rogue waves attracted a lot of attention, owing to their strange properties
\cite{solli_nature_2007,kibler_np_2010,montina_prl_2009,birkholz_prl_2013}.
As a pheonomenon, rogue wave originated in oceans, as an extreme localized wave that suddenly
appears and disappears without a trace. However, it
is now accepted that, as a {\it nonlinear} phenomenon,
it can be well described by the cubic nonlinear Schr\"odinger equation (NLSE).
The cubic NLSE possesses a variety of solutions \cite{kaminer_prl_2011, zhang_ol_2013, zhong.pre.87.065201.2013, zhangyiqi_oe_2014, zhong.pre.90.043201.2014},
some of which can serve as prototypical rogue waves.
These include Peregrine solitons \cite{peregrine_jams_1983,kibler_np_2010},
Kuznetsov-Ma breathers \cite{ma_sam_1979,kibler_sc_2012},
Akhmediev breathers (ABs) \cite{akhmediev_tmp_1987},
higher-order rogue waves \cite{erkintalo_prl_2011, kedziora_pre_2013},
and Fermi-Pasta-Ulam (FPU) recurrent pulses \cite{simaeys_prl_2001,mussot_prx_2014,wabnitz_pla_2014}.
It should be emphasized that these solutions are {\em not} rogue waves {\em per se} but can be used to model ones.
The true rogue waves are extreme wave phenomena that sporadically appear in the interaction of solutions of different NLSEs
and require statistical description for evidence and confirmation.
The solutions mentioned above are by and large exact solutions to NLSE that contain solitary or
trains of pulses, ride on finite backgrounds, and are prone to modulation instabilites.
It is in the devlopment of unstable wavefronts that extreme or freak waves may appear.

On the other hand, Talbot effect represents a self-imaging phenomenon in the near-field diffraction of plane waves,
first observed by H. F. Talbot \cite{talbot_1836}
and theoretically explained by Lord Rayleigh \cite{rayleigh_1881}.
This effect is a spatial recurrence phenomenon: if one records directly beyond a
narrow diffraction slit the beam intensity along the propagation direction,
one sees a series of periodic diminishings and revivals,
generated by the interference of diffracted secondary wavelets coming from different slits.
Stretched in the transverse dimension, one observes the so-called Talbot carpet.
The greatness of the effect lies in the fact that essentially all one needs to observe periodicity along the longitudinal direction is the periodicity of input in the transverse direction.

Owing to its potential applications in image preprocessing and synthesis, photolithography,
optical testing, optical metrology, spectrometry, and optical computing,
Talbot effect has been reported in, but not confined to,
atomic optics \cite{wen_apl_2011,yiqi_ieee_2012},
quantum optics \cite{song_prl_2011,jin_apl_2012},
waveguide arrays \cite{iwanow_prl_2005},
photonic lattices \cite{ramezani_prl_2012},
Bose-Einstein condensates \cite{deng_prl_1999,ryu_prl_2006},
X-ray imaging \cite{pfeiffer_nm_2008},
and in an interferometer for C$_{70}$ fullerene molecules \cite{brezger_prl_2002}.
Recent research indicates that Talbot effect can also be obtained using spherical waves \cite{azana.prl.112.213902.2014}.
It is worth mentioning that the aforementioned Talbot effect represents various linear cases \cite{winthrop.josa.4.373.1965, berry.jmo.43.2139.1996, wen_aop_2013},
even the so-called ``nonlinear effect'' from Ref. \cite{zhang_prl_2010}.
Although the nonlinear Talbot effect was apparently previously investigated \cite{zhang_prl_2010,wen_josab_2011,zhangyiqi_pre_2014},
it is different from the one that will be described here.
The topic is still in need of further exploration,
because interdisciplinary research -- in this case of rogue waves, Talbot effect, and
self-Fourier transformation -- tends to induce new ideas and applications.

Therefore, we will restrict our attention to just one basic rogue wave solution that
displays the nonlinear Talbot effect -- the Akhmediev breather. However, we will consider other inputs as well.
An AB wave is periodic along the transverse coordinate and decays along the longitudinal
coordinate -- it is a transverse train of single optical pulses.
When such a train is launched into a nonlinear Kerr medium, owing to modulation instability
and nonlinear interference of propagating pulses, a Talbot recurrence phenomenon is observed.
Hence, if the medium is nonlinear, then one obtains the nonlinear Talbot effect.
There exists a similar nonlinear recurrence phenomenon in fibers -- the Fermi-Pasta-Ulam recurrence;
it's 60th birthday was just celebrated
\cite{mussot_prx_2014,akhmediev_nature_2001,simaeys_prl_2001,simaeys_josab_2002}.
Nevertheless, the two effects are different in nature.
Even though both can be described by the NLSE,
FPU is a temporal phenomenon, whereas the nonlinear Talbot effect is a spatial phenomenon.
The nonlinear Talbot effect of rogue waves has recently been reported \cite{zhangyiqi_pre_2014}.
Although a periodic initial condition is always required for a Talbot effect,
the nonlinear Talbot effect is in stark contrast to the linear one.
The linear Talbot requires real gratings or periodic diffracting structures for generation,
it forms in linear homogenous media, and can be generally explained by the Fresnel diffraction theory \cite{wen_aop_2013}.
In contrast, the nonlinear Talbot effect
requires periodic inputs that ride on a finite background, happens in bulk NL media, and possesses no adequate theoretical explanation.

In this paper, we report both linear and nonlinear Talbot effects in two dimensions, but
focus on the nonlinear Talbot effect formed from rogue and other waves in a bulk nonlinear medium.
The two-dimensional nonlinear Talbot effect is investigated here for the first time, to the best of our knowledge.
The two-dimensional (2D) Talbot effect is essentially a 3D optical phenomenon;
it results in the formation of 3D periodic optical structures.
In passing, we also note that Talbot effect can be regarded as a self-Fourier transform of input beams.
We first construct the incidence from a product of two 1D (doubly periodic) ABs,
and then propagate it in a linear and nonlinear Kerr medium.
The result is a 3D stack of Talbot carpets.
Similar to the nonlinear 1D Talbot effect reported in Ref. \cite{zhangyiqi_pre_2014}, which not only originates from an exact rogue wave but also
requires bulk NL medium to form,
such 1D and 2D nonlinear Talbot effects are different from those reported in Ref. \cite{zhang_prl_2010}.

The organization of the paper is as follows.
In Sec. \ref{model} we briefly introduce the theoretical model and construct the input from rogue wave solutions.
In Sec. \ref{linearTalbot} linear Talbot effect (coming also from rogue waves) is reviewed theoretically and numerically,
as a background for the nonlinear Talbot effect.
In Sec. \ref{nonlinearTalbot} we numerically investigate in some detail the nonlinear Talbot effect from rogue wave solutions,
as well as from other periodic inputs.
The paper is concluded with Sec. \ref{conclusion}.

\section{The Model}\label{model}

The propagation of a beam with envelope $\psi$ in a Kerr medium
can be generally described by the normalized cubic NLSE
\begin{equation}\label{nls1}
  i\frac{\partial \psi}{\partial z} + \frac{1}{2} \nabla_\bot^2 \psi + |\psi|^2 \psi = 0,
\end{equation}
where $\nabla_\bot^2$ is the transverse Laplacian; in one dimension it is $\nabla_\bot^2=\partial^2/\partial x^2$
and in two dimensions $\nabla_\bot^2=\partial^2/\partial x^2 + \partial^2/\partial y^2$.
The transverse coordinates are measured in units of some characteristic transverse length,
whereas the longitudinal coordinate is given in terms of the corresponding Rayleigh range.
One of the rogue solutions of the 1D NLSE is the Akhmediev breather \cite{akhmediev_tmp_1987}
\begin{align}\label{akh1}
  \psi(z,\,x) =& \left[ \frac{(1-4q_x)\cosh(a_x z) + \sqrt{2q_x} \cos(\Omega_x x) }{\sqrt{2q_x}\cos(\Omega_x x) - \cosh(a_x z)} + 
   i\frac{a \sinh(a_x z)}{\sqrt{2q_x}\cos(\Omega_x x) - \cosh(a_x z)} \right] \times \exp(iz),
\end{align}

where  \[a_x=\sqrt{8q_x(1-2q_x)}\] and  \[\Omega_x=2\sqrt{1-2q_x},\] with $q_x$ being a free parameter ranging from 0 to 1/2.
This solution is periodic in the transverse direction
and dies fast in the propagation ($z$) direction.
The period of $\psi(z,\,x)$ along $x$ axis is \[\mathcal{D}_x=\frac{\pi}{\sqrt{1-2q_x}}\] and
the maximum of $|\psi(z,\,x)|^2$ is \[M_x=\left|\frac{1+\sqrt{2q_x}-4q_x}{1-\sqrt{2q_x}}\right|^2.\]
We have previously demonstrated that a beam with the profile of AB at $z=0$
[viz. $\psi(z=0,\,x)$] will not die away when propagated
according to the 1D NLSE, but thanks to modulation instability and
nonlinear interference it will display the nonlinear Talbot effect \cite{zhangyiqi_pre_2014}.
A question naturally arises, can this be generalized to 2D?

Even though various rogue wave solutions are reported for the 1D NLSE \cite{kedziora_pre_2013},
unfortunately there exist no analytical breather or periodic solutions of the 2D NLSE.
It would be nice if such an eigenfunction of 2D NLSE exists in the transverse plane;
then it would be possible to check directly if it produces the 2D nonlinear Talbot effect.
Since this is not the case, the next best possibility is to use the product of two 1D ABs as a planar diffraction pattern, to see
if it produces the effect.
This product is not the solution of 2D NLSE, nonetheless it is useful in investigating the nonlinear Talbot effect.
In a latter section, we discuss the possibility of obtaining nonlinear Talbot effect from simpler periodic 2D inputs.
Thus, the incident wave for the 2D NLSE is chosen in the form
\begin{align}\label{akh2}
  \psi(x,\,y) = \frac{C}{\sqrt{M_xM_y}}   \frac{(1-4q_x) + \sqrt{2q_x} \cos(\Omega_x x)}{\sqrt{2q_x}\cos(\Omega_x x) - 1} 
   \frac{(1-4q_y) + \sqrt{2q_y} \cos(\Omega_y y)}{\sqrt{2q_y}\cos(\Omega_y y) - 1},
\end{align}
which is the product of two perpendicular 1D ABs $\psi(z=0,\,x)$ and $\psi(z=0,\,y)$.
It is clear that $\psi(x,\,y)$ is periodic both along $x$ and $y$ cooridinates.
The periods along $x$ and $y$ directions are $\mathcal{D}_x$ and $\mathcal{D}_y$, respectively.
Generally, $\mathcal{D}_x\neq \mathcal{D}_y$, but if $q_x=q_y$, $\mathcal{D}_x=\mathcal{D}_y$.
The coefficient $1/\sqrt{M_xM_y}$ in Eq. (\ref{akh2}) makes the maximum amplitude equal to that of Eq. (\ref{akh1}), if $C=1$.

Note that the initial beam can be formed by other periodic functions in the transverse plane, not only by the product of two ABs.
However, we believe that not all periodic functions can be used to produce nonlinear Talbot effect.
We choose the product of two ABs with $z=0$, because AB is the solution of 1D NLSE
and exhibits convenient properties in comparison with other solutions \cite{kaminer_prl_2011,zhang_ol_2013,zhangyiqi_oe_2014},
that is, it is periodic along both transverse coordinates and has a constant nonzero background.
Another solution of the cubic NLSE that possesses similar properties is the doubly-periodic AB \cite{akhmediev_tmp_1987}
\begin{align}\label{eq6}
  \psi(z,\,x)= k_x \frac{A(x) \, {\rm dn}\left( k_x z,\,{1}/{k_x} \right) + i/{k_x} \,{\rm sn} \left( k_x z,\,{1}/{k_x} \right)}
  {\left[ 1-A(x) \, {\rm cn} \left( k_x z,\,{1}/{k_x} \right) \right]} 
\exp(iz),
\end{align}
in which ${\rm sn}$, ${\rm cn}$ and ${\rm dn}$ stand for Jacobian elliptic functions,
and \[A(x)=\sqrt{{1}/(1+k_x)} \, {\rm cn} \left( \sqrt{2k_x}x,\, \sqrt{(k_x-1)/{2k_x}} \right)\] with $k_x>1$.
Because ${\rm sn}(x,m)$, ${\rm cn}(x,m)$ and ${\rm dn}(x,m)$ are all periodic and the corresponding periods are $4K$, $4K$ and $2K$, respectively,
with \cite{abramowitz_book} \[K =\int_0^{\pi/{2}}\frac{1}{\sqrt {1-m^2\sin^2\theta}} {d\theta},\]
the solution described by Eq. (\ref{eq6}) is periodic along both
$x$ and $z$ directions.
The transverse period is
\[\mathcal{D}_x=\frac{4}{\sqrt{2k}}K_{m=\sqrt{(k_x-1)/(2k_x)}}.\]
As a result, one can also construct a viable initial 2D beam by using the product of two doubly-periodic ABs at $z=0$.

In the following sections,
we study the propagation and dynamics of the product of two rogue waves,
in both linear and nonlinear bulk media. We also consider Talbot effect coming from simpler periodic inputs.
To guarantee high numerical precision, we utilize a fourth-order split-step fast Fourier transform (FFT)
method \cite{yang_book} in double precision.
To make beams of finite energy and prevent FFT spillover effects,
we use an aperture with a diameter large enough to enforce fast
convergence of beam intensity to zero at the transverse infinity.

\section{Linear Talbot Effect}\label{linearTalbot}

\subsection{Theoretical analysis}

We discuss first the linear propagation equation [that is, the nonlinear term in Eq. (\ref{nls1}) is eliminated].
Using the separation of variables method, one ends up with the following coupled equations \cite{yang.ctp.53.937.2010}:
\begin{subequations}
\begin{equation}\label{ceq1}
  i\frac{\partial}{\partial z} \psi(x,\,z) + \frac{1}{2} \frac{\partial^2}{\partial x^2} \psi(x,\,z) - \mu \psi(x,\,z) = 0,
\end{equation}
\begin{equation}\label{ceq2}
  i\frac{\partial}{\partial z} \psi(y,\,z) + \frac{1}{2} \frac{\partial^2}{\partial y^2} \psi(y,\,z) + \mu \psi(y,\,z) = 0,
\end{equation}
\end{subequations}
where $\mu$ is the separation constant.
If we set $\psi(x,\,z)=f(x,\,z)\exp(-i\mu z)$ and $\psi(y,\,z)=g(y,\,z)\exp(i\mu z)$,
Eqs. (\ref{ceq1}) and (\ref{ceq2}) can be recast into
\begin{subequations}
\begin{equation}\label{ceq3}
  i\frac{\partial}{\partial z} f(x,\,z) + \frac{1}{2} \frac{\partial^2}{\partial x^2} f(x,\,z) = 0,
\end{equation}
\begin{equation}\label{ceq4}
  i\frac{\partial}{\partial z} g(y,\,z) + \frac{1}{2} \frac{\partial^2}{\partial y^2} g(y,\,z) = 0,
\end{equation}
\end{subequations}
which are the same as the 1D case investigated in Ref. \cite{zhangyiqi_pre_2014};
the initial fields are $f(x,\,z=0)=\psi(x,\,z=0)$ and $g(y,\,z=0)=\psi(y,\,z=0)$, respectively.
Therefore, the 2D problem is reduced to the two independent 1D problems,
which greatly decreases the complexity of the problem.
The solutions of Eqs. (\ref{ceq3}) and (\ref{ceq4}) have the same format, which can be written as \cite{bernardini.epj.16.58.1995}
\begin{align}\label{solution}
  \vartheta(x,\,z) =  \sqrt{-\frac{i}{2\pi z}} \exp\left(i\frac{x^2}{2z}\right) 
 \int_{-\infty}^{+\infty} d\xi \left[ \vartheta(\xi,\,0) \exp\left(i\frac{\xi^2}{2z}\right) \right] \exp\left(-i\frac{x}{z}\xi\right),
\end{align}
where $\vartheta$ stands $f$ or $g$.
It is clear that Eq. (\ref{solution}) describes the Fresnel diffraction \cite{goodman.book}.
Therefore, one can obtain the analytical expressions of the Talbot length corresponding to the two components \cite{yiqi_ieee_2012},
\begin{align}\label{tl}
  z_{Tx}=\frac{\mathcal{D}_x^2}{\pi},\quad z_{Ty}=\frac{\mathcal{D}_y^2}{\pi}.
\end{align}
Since the 2D linear Talbot effect can be treated as a product of two independent 1D linear Talbot effects,
the Talbot length of the 2D linear Talbot effect would be the \textit{least common multiple} of two 1D Talbot lengths.

In addition, based on Eq. (\ref{solution}), the initial periodic beam can be arbitrary.
Therefore, we can investigate the linear Talbot effect from a product of two ABs, two doubly-periodic ABs,
or any other kind of periodic beams.
In light of the formation being quite similar, for the linear case
we only focus on the product of two ABs with equal periods.
For convenience, we set $q_x=q_y=q$, $\mathcal{D}_x=\mathcal{D}_y=\mathcal{D}$, and $M_x=M_y=M$.

\subsection{Numerical simulation}

\begin{figure}[htbp]
\centering
  \includegraphics[width=0.5\columnwidth]{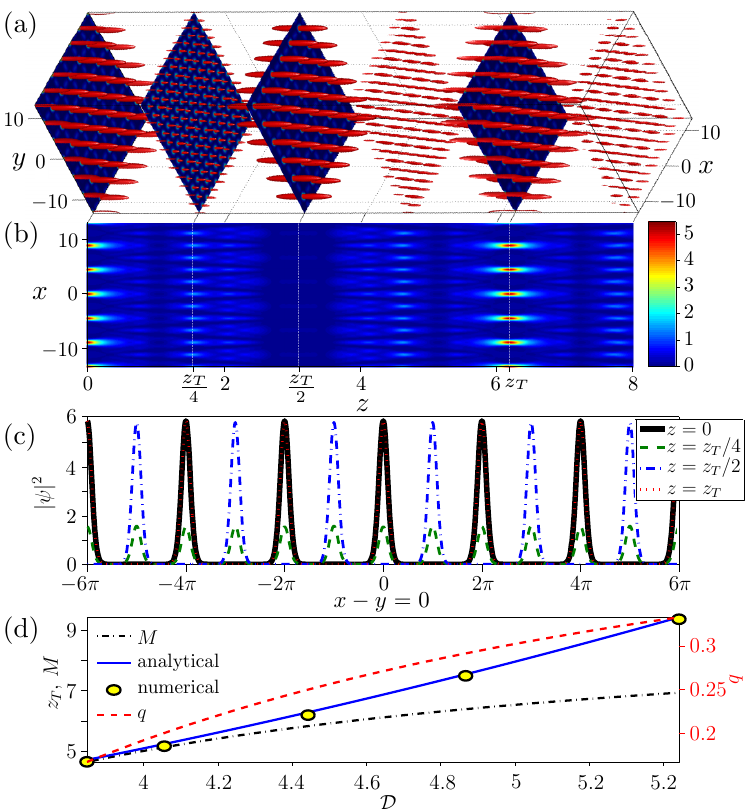}
  \caption{(Color online) Linear 2D Talbot effect.
  (a) Intensity isosurface plot of the propagating rogue wave incidence, according to the linear Schr\"odinger equation.
  The beam intensities are displayed at the distances of $1/4$, $1/2$ and 1 Talbot lengths.
  (b) Corresponding intensity carpet in the $xz$ plane at $y=0$.
  (c) Intensity profiles along the $x/y$ axes, depicting the profiles at $z=0$, $z=z_T/4$, $z=z_T/2$ and $z=z_T$ from (a).
  Other parameters: $q=1/4$, $\mathcal{D}=\sqrt{2}\pi$ and $M=3+2\sqrt{2}$.
  (d) Relationships between $\mathcal{D}$ and $z_T$, $q$, and $M$.}
  \label{fig1}
\end{figure}

In Fig. \ref{fig1}(a) we display the transverse intensity isosurface distribution of the propagating beam with $q=1/4$,
at different propagation distances. It is immediately seen that
the incident rogue wave reappears at $z\approx6.192$.
This distance defines the Talbot length $z_T$.
At $z_T/4$ and $z_T/2$, the fractional Talbot images also form.
Similar to previous research \cite{iwanow_prl_2005,zhang_prl_2010,yiqi_ieee_2012},
the fractional Talbot image at $z_T/2$ exhibits a $\pi$ phase shift in comparison with the incidence,
while for the $z_T/4$ fractional Talbot image, the transverse period is halved.
To see the Talbot effect more clearly,
we also plot the intensity carpet versus $x$ and $z$ in the plane $y=0$, in Fig. \ref{fig1}(b).
The Talbot images at $z_T/4$ and $z_T$ are quite apparent, while the one at $z_T/2$ is missing.
The reason is that because of the $\pi$ phase shift, the Talbot image at $z_T/2$ is shifted for half of the period transversely
and cannot be seen in the $y=0$ plane.
Supplementary material \cite{Note1} provides a clear animated 2D version of this propagation.
In Fig. \ref{fig1}(c), the intensity profiles at $z=0$, $z=z_T/4$, $z=z_T/2$, and $z=z_T$ along the $x/y$ axes are displayed.
In conclusion, the 2D linear Talbot effect of rogue waves is easily visible and verified.

It is clear that the linear Talbot effect does not change with different incident intensities.
Therefore, one can check the influence of $q$ (also of the transverse period) on the formation of linear Talbot effect,
even though the intensity of the incident beam changes with $q$ \cite{zhangyiqi_pre_2014}.
In Fig. \ref{fig1}(d), we display the relation between $\mathcal{D}$ and Talbot length, as shown by the
solid curve and circles, which correspond to the analytical and numerical results, respectively.
It is seen that these results agree with each other quite well.
The same figure also displays
the relationship of $\mathcal{D}$ vs $q$ (dashed curve) and $\mathcal{D}$ vs $M$ (dash-dotted curve).

\subsection{Talbot effect as a self-Fourier transform}

For a moment, let us go back to Eq. (\ref{solution}).
It is evident that the integral can be viewed as a Fourier transform of $\vartheta(x,\,0) \exp\left(ix^2/2z\right)$
with the spatial frequency $x/z$.
At the Talbot length $z_T$, the initial beam is reproduced \cite{lohmann.josaa.9.2009.1992},
i.e., $\vartheta(x,\,z_T) = \vartheta(x,\,0)$
which means that the Fourier transform of $\vartheta(x,\,0) \exp\left(ix^2/2z_T\right)$ with the frequency $x/z_T$
is the initial beam $\vartheta(x,\,0)$.
This result extends to the two-dimensional case straightforwardly.
Therefore, the Talbot effect can be viewed as a \textit{self-Fourier transform} of input periodic beams
that are periodically reconstructed during propagation.

At an arbitrary $z$,
Eq. (\ref{solution}) can be scaled through variable substitution $x' = x\sqrt{z_T/z}$ and $\xi' = \xi\sqrt{z_T/z}$, as
\begin{align}\label{solution2} 
  \vartheta(x,\,z) = \sqrt{-\frac{i}{2\pi z_T}} \exp\left(i\frac{x'^2}{2z_T}\right) 
 \int_{-\infty}^{+\infty} d\xi' \left[ \vartheta\left(\sqrt{\frac{z}{z_T}}\xi',\,0\right) \exp\left(i\frac{\xi'^2}{2z_T}\right) \right] \exp\left(-i\frac{x'}{z_T}\xi'\right).
\end{align}
As a result, the fractional Talbot effect is the self-Fourier transform of the scaled transverse initial periodic beam.
It is worth mentioning that the scaling is equivalent to a $\pi$ phase shift at $z=z_T/2$, due to the periodicity of the initial beam.
In fact, the linear Talbot effect can be interpreted as a {\it fractional} self-Fourier transform \cite{lohmann.josaa.9.2009.1992}.
However, the nonlinear Talbot effect cannot be described as such, because it
does not contain fractional images.
The nonlinear Talbot effect can be classified only as a regular self-Fourier transform,
in the sense
that it only contains the original image at $z_T$ and the shifted original image at $z_T/2$.
This question will be addressed in the following section.

\section{Nonlinear Talbot effect}\label{nonlinearTalbot}

When the nonlinearity is included, the practical problem is that the model cannot be treated by the method of separation of variables.
Nonetheless,
following the same numerical procedure as in the linear case, the 2D nonlinear Talbot effect
can be investigated and demonstrated.
We first take the product of two ABs with equal transverse periods as the initial beam,
then a product of two doubly-periodic ABs,
and at last a relatively simple periodic function.

\subsection{Akhmediev breather as an input}

\begin{figure}[htbp]
\centering
  \includegraphics[width=0.5\columnwidth]{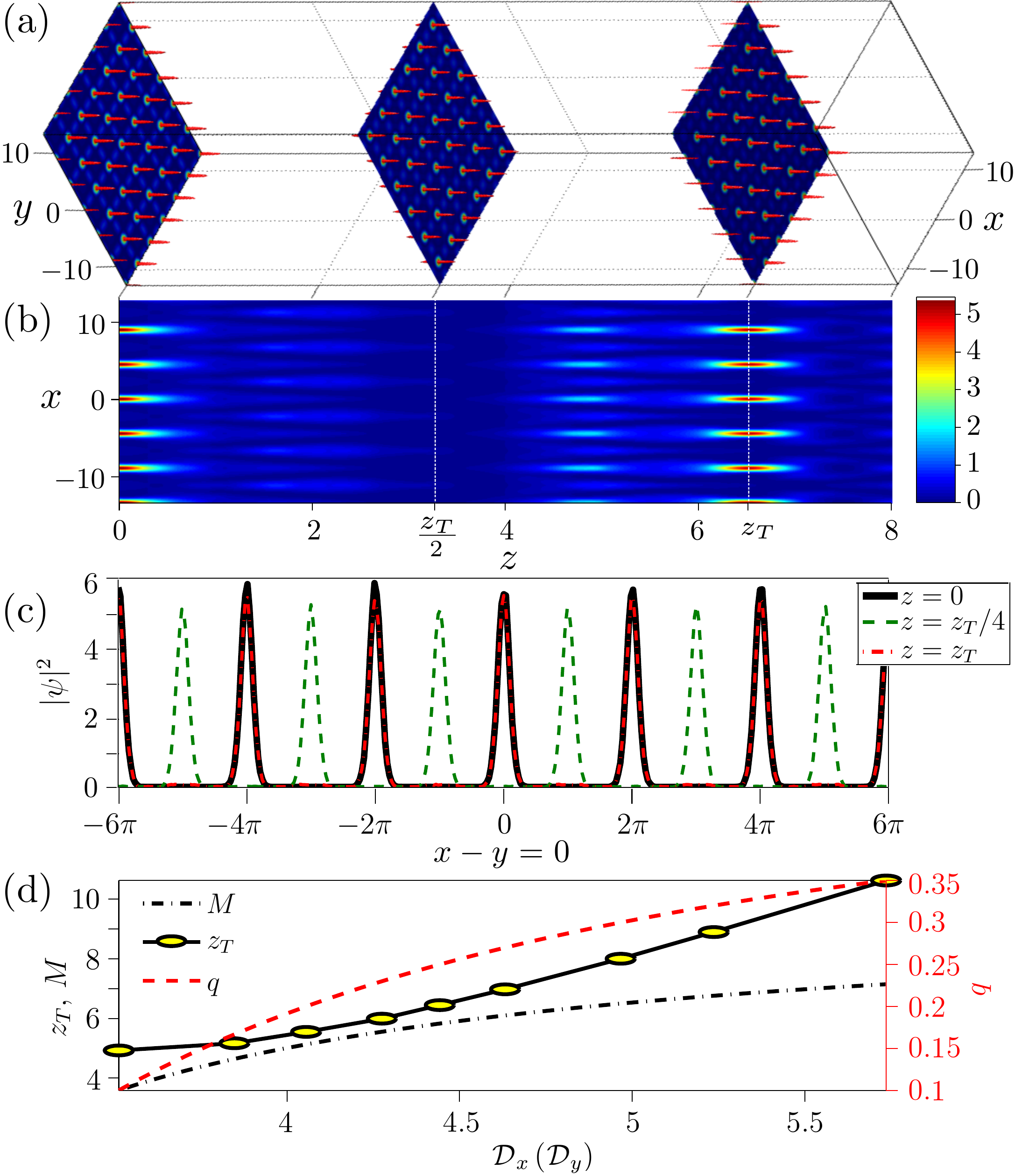}
  \caption{(Color online) Same as Fig. \ref{fig1}, but for the 2D nonlinear Talbot effect.}
  \label{fig2}
\end{figure}

Results for a product of ABs are presented in Fig. \ref{fig2}, which follows the same setup as Fig. \ref{fig1},
but with a perturbation added to the input beam.
The maximum amplitude of the perturbation is $10\%$ of the product of ABs.
The perturbation is added to check the stability and robustness of the beams in propagation,
which is necessary for nonlinear propagation.
In Fig. \ref{fig2}(a) we show the isosurface intensity plots that display the formation of 2D nonlinear Talbot effect at $q=1/4$.
Together with the isosurface plot, the intensities at certain distances are also shown.
One striking difference with the linear Talbot effect is immediately apparent:
the fractional Talbot images are gone. One can only observe the secondary and primary Talbot images at $z_T/2$ and $z_T$.
This is fundamentally different from the 2D linear Talbot effect.

We also display the intensity carpet in the $y=0$ plane in Fig. \ref{fig2}(b),
from which one can verify that there are no fractional Talbot images.
It is evident that in the propagation range $z\in[z_T/2,~z_T]$ a structure similar to the fractional Talbot image appears,
but this is not a fractional Talbot image, because of the following reasons:
(1) The transverse period is the same as that of the input; (2) The position is not at $3z_T/4$.
These structures are just the consequence of modulation instability.
Such images also exist in the range $z\in[0,~z_T/2]$, however one cannot see them in Fig. \ref{fig2}(b)
because the maximum is shifted from the plane $y=0$, similar to the image at $z=z_T/2$.
More clear details about the nonlinear, as well as linear evolution of Talbot images,
can be seen in the movies provided in the Supplemental Material
\footnote{See Supplemental Material at \url{http://link.aps.org/supplemental/}
for the animated version of the propagations of the beams shown in Figs. \ref{fig1}(a) and \ref{fig2}(a).}.

The intensity profiles of the beam at $z=0$, $z=z_T/2$ and $z=z_T$ along the $x/y$ axes are shown in Fig. \ref{fig2}(c).
Even though the intensity carpet in Fig. \ref{fig2}(b) is not entirely symmetric about $z=z_T/2$,
the image at $z=z_T$ is the same as the input, and that at $z_T/2$ has a $\pi$ phase shift (and is absent from the figure).
The intensity profiles in Fig. \ref{fig2}(c) demonstrate the formation of the 2D nonlinear Talbot effect.
Note that the modulation instability in the Kerr nonlinear medium results in the formation of
a 2D nonlinear Talbot carpet that is not perfect -- the intensity maximum is a bit smaller than that of the incidence
and small humps appear between the neighboring peaks.
However, the formation of 2D nonlinear Talbot effect of rogue waves in bulk nonlinear medium is clearly demonstrated,
because the required images at $z_T/2$ and $z_T$ appear.
Again, one should check the evolution displayed in the movie in the Supplementary Material \cite{Note1},
to ascertain the formation of 2D nonlinear Talbot effect more convincingly.

Since $q$ (or the transverse period) is related to the intensity of the input beam
and optical response of the nonlinear medium is sensitive to the beam intensity,
the formation of 2D nonlinear Talbot effect with different transverse periods will be much more complex than that of the 2D linear Talbot effect.
In Fig. \ref{fig2}(d) we show the changing trend of the Talbot length versus the transverse period.
The relationships between the transverse period and $q$ as well as $M$ are also displayed,
which are similar to those in Fig. \ref{fig1}(d).
Phenomenologically, the Talbot length increases slowly with the increase in $D_x$ for $q<0.2$,
while for $q>0.2$ the relation is more steep.
Viewed as a whole, the relation between $D_x$ and $z_T$ is parabolic.
This phenomenon can be explained from a scaling point of view:
rescaling transverse coordinates by a factor scales the longitudinal coordinate by the factor squared.
In the linear case the relation can also be explained by the same scaling law, as depicted by Eq. (\ref{tl}).

Similar to the linear Talbot effect, the
nonlinear Talbot effect can be also viewed as a self-Fourier transform.
At $z=z_T$, one obtains the initial beam, while at $z=z_T/2$ one obtains the same beam but transversely shifted.
At other places, the ``Fourier transform'' still gives a periodic image,
but it is different from the initial beam even when the transverse scaling is taken into account.
Hence, it cannot be viewed as the fractional Talbot effect. The nonlinear Talbot effect proceeds from an initial
beam to the Fourier transform at $z=z_T/2$ (which is the shifted beam) and back to the original beam
at $z=z_T$. Thus, it can be described as a genuine self-Fourier transform.

\subsection{Discussion}

The amplitude of the incident 2D AB in Eq. (\ref{akh2}) can be adjusted by changing the value of $C$,
so we investigate numerically the relationship between the intensity and Talbot length for the same transverse period.
This is important because of the potential wave collapse in 2D.

\begin{figure}[htbp]
\centering
  \includegraphics[width=0.5\columnwidth]{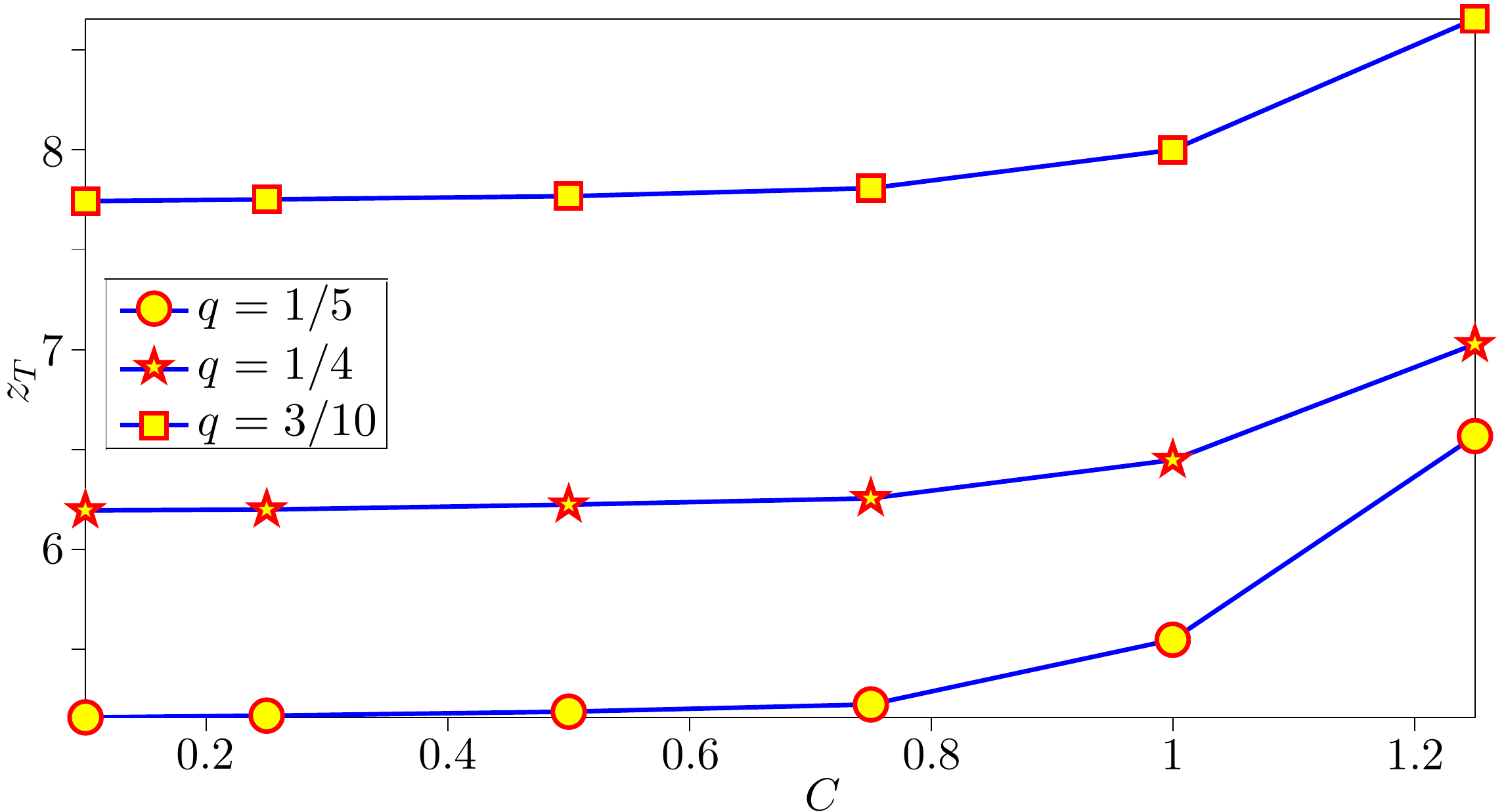}
  \caption{(Color online) Talbot dependence on the amplitude of the incident with fixed transverse periods.}
  \label{fig3}
\end{figure}

We fix $q$ (by correspondingly fixing $D_x$ and $D_y$) and change the amplitude of the incident beam, to calculate the Talbot length;
the results are depicted in Fig. \ref{fig3}.
Three values of $q$ are chosen, $1/5$, $1/4$ and $3/10$, respectively.
As shown in Fig. \ref{fig2}(d), the bigger the $q$, the larger the Talbot length $z_T$,
so that in Fig. \ref{fig3}, the Talbot length is the biggest when $q=3/10$, for the same $C$.
With $C$ increasing, the Talbot length also increases,
and such increment becomes quite fast when $C>1$.
The reason is that higher intensity leads to stronger modulation instability,
which demands a longer distance to adjust itself during propagation.
Since the nonlinearity is Kerr in NLSE, the value of $C$ cannot be chosen very high.
Numerical simulations indicate that the propagation will collapse in a short distance
when $C=2$, even with a small perturbation.
However, we would like to emphasize that the beam will always collapse in such a medium
when the propagation distance is long enough; the
occurrence of collapse does not depend of the value of $C$ only, but on the value of $q$ as well.
Choosing proper $C$ can only help us demonstrate nonlinear Talbot effect over a relatively long distance
-- in Fig. \ref{fig2} the beam could robustly propagate over 8 Rayleigh lengths. Finite
longitudinal extension of the Talbot effect is also a consequence of the finite transverse window over which
the propagation is considered. The wider the window -- the longer the carpet. The longitudinal periodicity
of Talbot images is the consequence of the transverse periodicity of input beams.

Finally, we touch upon the touchy subject of wave collapse.
It is clear that the well-known ``catastrophic self-focusing collapse'' may occur in cubic nonlinear medium
during the development of nonlinear Talbot effect,
i.e., the beam will always collapse during propagation if the input power is high enough and the propagation distance long enough.
Therefore, to obtain a relatively stable recurrence, one should use not too high input intensity.
An interesting topic would be to address the questions of threshold intensity and generally the stability of nonlinear
Talbot effect when different parameters in the model are varied. These questions, however are beyond the scope of this paper.
To check for an eventual development of collapse, we only added a small amplitude perturbation to the initial beam.
As mentioned, in numerical simulations we introduce white noise,
whose maximum amplitude is up to $10\%$ of the input beam. We find that the
figures reported in this paper remain the same.

\subsection{Results from the doubly periodic ABs}

\begin{figure} [htbp]
	\centering
	\includegraphics[width=0.5\columnwidth]{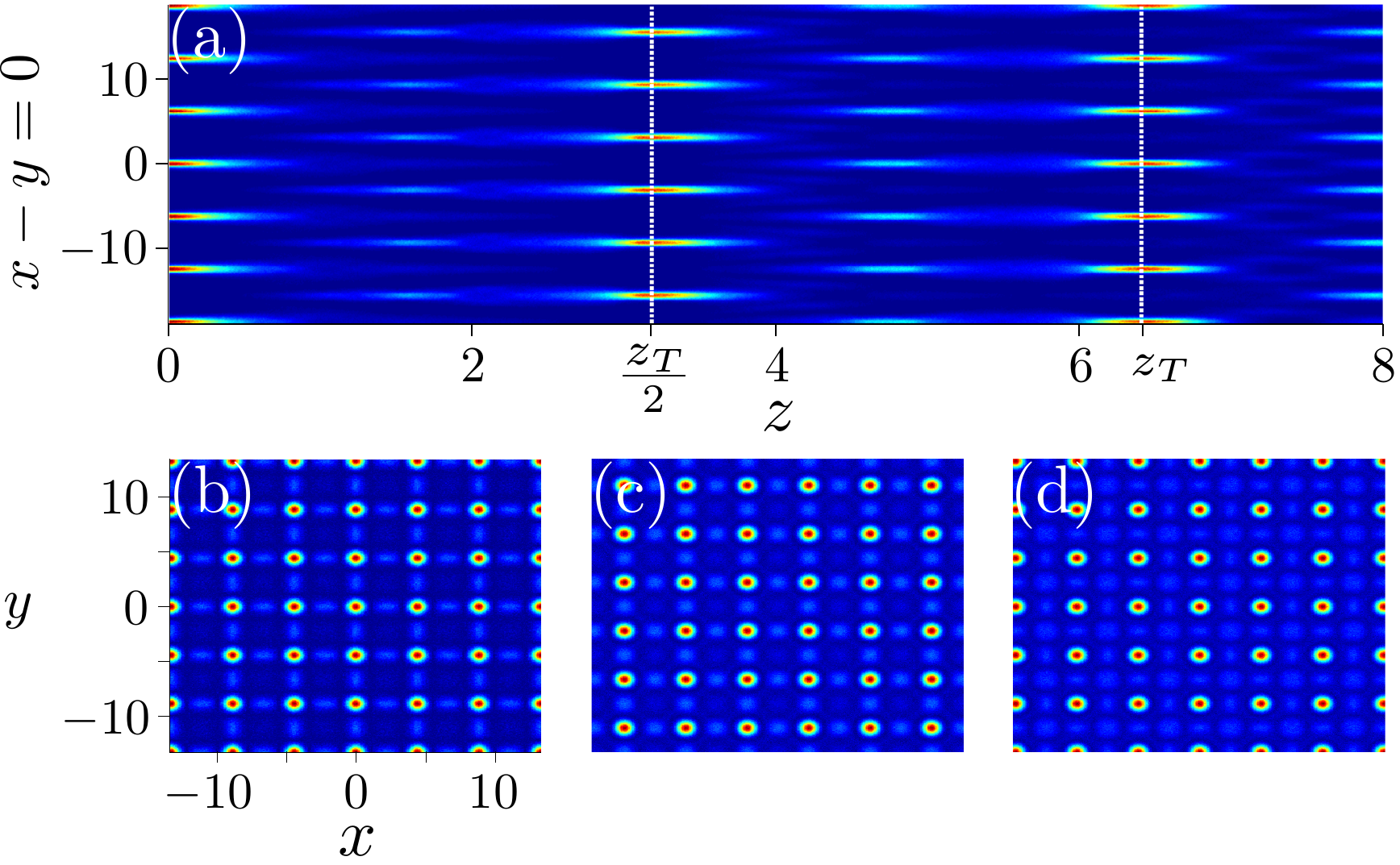}
	\caption{(Color online)
		(a) 2D Talbot effect coming from a product of two doubly periodic ABs with $k=1.2$ in the plane $x-y=0$.
		Dashed lines show the locations of half Talbot length and Talbot length, respectively.
		(b)-(d) Intensities at the initial place, half Talbot length and Talbot length.}
	\label{fig4}
\end{figure}
An input in the form of a product of two orthogonal doubly periodic ABs is obtained
using Eq. (\ref{eq6}) and setting $z=0$.
In the execution of numerical simulations, a perturbation is also introduced.
Results are displayed in Fig. \ref{fig4}. The formation of the 2D Talbot effect from a product of two doubly periodic ABs is evident.
In Fig. \ref{fig4}(a), we display the intensity evolution in the plane $x-y=0$,
so as to depict the generation of the Talbot carpet clearly.

We can see that the beam is robust in propagation over such a long distance, even though the perturbation is included.
Numerically, we find that the Talbot length is about $z_T\approx6.436$,
as shown by the right dashed line in Fig. \ref{fig4}(a).
In addition, we also show the beam intensities at $z=0$, $z=z_T/2$, and $z=z_T$ in Figs. \ref{fig4}(b)-\ref{fig4}(d).
Therefore, it is feasible to obtain nonlinear Talbot effect from a product of two doubly periodic ABs.
The question is, can it be obtained from more simple periodic patterns.

\subsection{Results from simply periodic initial conditions}

In the above sections, we were mainly concerned with the incident beams constructed from rogue waves.
What about the simpler periodic initial conditions?
Is rogue wave a necessary condition?
To answer these questions,
we construct a transversely periodic and longitudinally localized beam as follows
\begin{equation}\label{cosine_input}
\psi(x,\,z) = 1 + \cos(2x) \exp\left(-z^2\right),
\end{equation}
which  possesses a nonzero background.
By making a product of $\psi(x,\,z=0)$ and $\psi(y,\,z=0)$,
one gets a 2D input beam in the form
\begin{equation}\label{cosine}
\psi(x,\,y) = C [1 + \cos(2x)][1 + \cos(2y)].
\end{equation}
Here, we assume $C=0.25$, to make the maximum of the input 1.

\begin{figure}[htbp]
\centering
  \includegraphics[width=0.5\columnwidth]{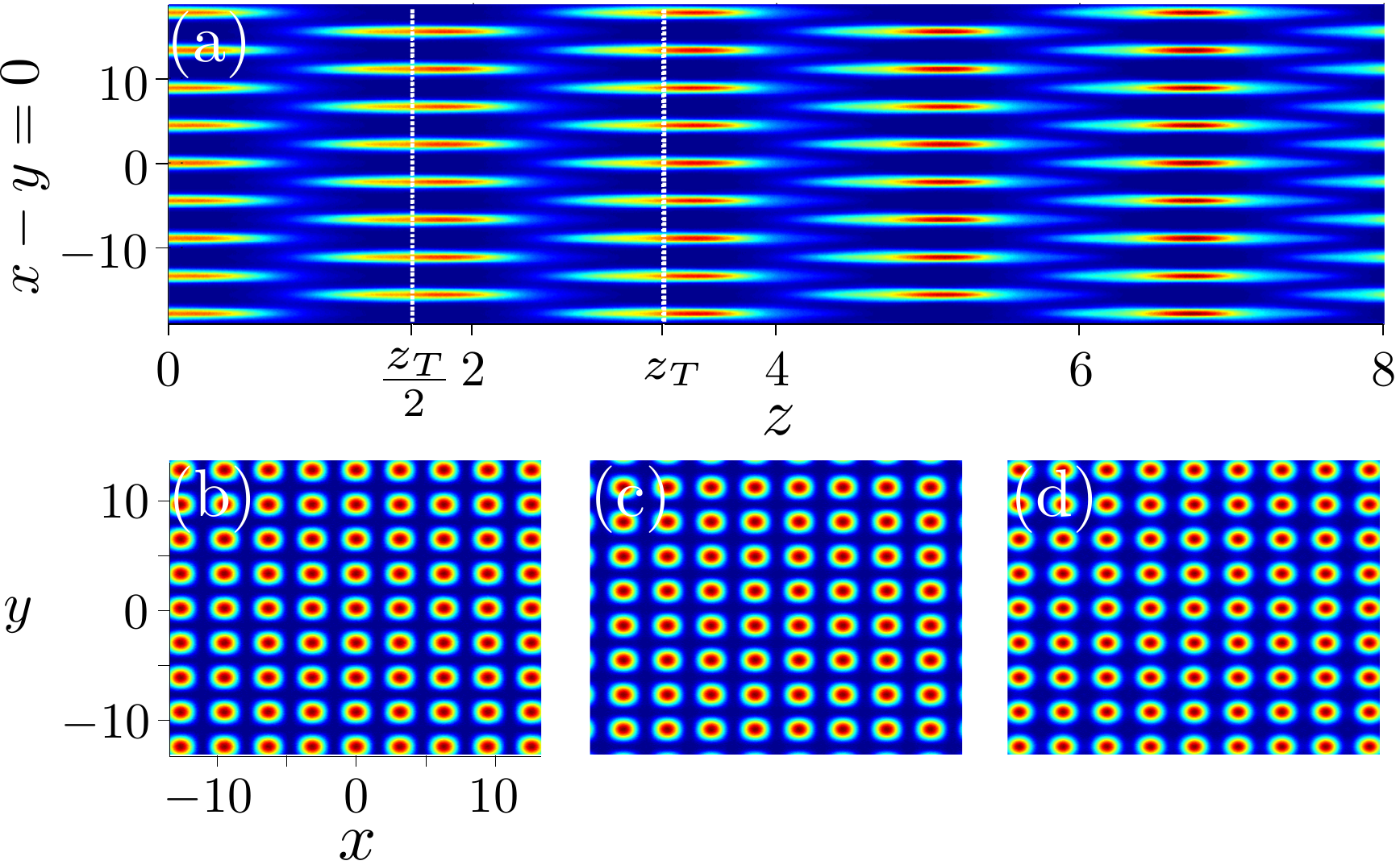}
  \caption{(Color online)
  2D Talbot effect resulting from Eq. (\ref{cosine}).
  Figure setup is as in Fig. \ref{fig4}.}
  \label{fig5}
\end{figure}

The propagation of the input (\ref{cosine}) with a small perturbation
(the maximum amplitude $\sim 5\%$ of the input)
is shown in Fig. \ref{fig5}.
From the figure one can see that the nonlinear Talbot recurrence forms during propagation (at $z_T\approx3.352$),
even though the peak intensity increases with the distance.
It is also seen that the nonlinear Talbot effect still does not exhibit the fractional Talbot images,
and that the evolution is similar to those shown in waveguide arrays \cite{iwanow_prl_2005}.
This demonstrates that a rogue wave solution is not a necessary condition to realize nonlinear Talbot effect.
In other words, nonlinear, as well as linear, Talbot effect is not limited to very specific initial conditions.

On the other hand, if $\psi$ in Eq. (\ref{cosine}) is chosen as a simple 2D pattern $\cos(2x) \cos(2y)$,
one can not realize the nonlinear Talbot effect (not shown here).
Therefore, one can say that not only rogue wave solution is not a necessary condition,
but also the initial condition can not be an arbitrary 2D periodic pattern.
The common condition is that the input should be constructed from periodic functions with nonzero background,
as displayed in Eqs. (\ref{akh1}), (\ref{eq6}) and (\ref{cosine_input}).

\section{Conclusion}\label{conclusion}

We have introduced the 2D nonlinear Talbot effect theoretically and displayed
the effect numerically, for the first time to the best of our knowledge.
For the effect to be seen, not only the incident beam has to be nonlinearly prepared,
but the formation of 2D Talbot images should also proceed in the bulk Kerr nonlinear medium.
We have also demonstrated the 2D linear Talbot effect in the form of a stack of Talbot carpets behind a
periodic 2D diffraction pattern. Thus, the 2D Talbot effect results in the formation of 3D periodic optical structures.
Different from the 2D linear Talbot effect, there are no fractional Talbot images in the 2D nonlinear Talbot effect.
Numerical experiments demonstrate that the smaller the transverse period
and the smaller the amplitude of the incident beam,
the shorter the Talbot length in 2D nonlinear Talbot effect.

In addition, numerical simulations also demonstrate that a rogue wave solution is not a necessary condition
for the realization of the 2D nonlinear Talbot effect. It may be realized from other, simpler periodic inputs,
but with a general requirement that they possess a finite background.
Last but not least, Talbot effect
can be viewed as self-Fourier transform of the initial periodic beams occurring during propagation.
We hope that our research has broadened up the potential applications of Talbot effect --
in particular, for possible fabrication of 3D photonic crystals --
and has deepened the understanding of rogue waves.

\section*{Aknowledgement}

This work was supported by the 973 Program (2012CB921804),
KSTIT of Shaanxi Province (2014KCT-10),
NSFC (11474228 and 61308015),
NSFC of Shaanxi province (2014JQ8341),
CPSF (2014T70923 and 2012M521773),
and the NPRP 6-021-1-005 project of the Qatar National Research Fund (a member of the Qatar Foundation).
Authors appreciate the anonymous referees for their constructive comments to improve the paper.

%

\end{document}